\magnification=1000 \vsize=25truecm \hsize=16truecm \baselineskip=0.6truecm \parindent=0truecm
\parskip=0.2cm
\nopagenumbers \font\scap=cmcsc10 \hfuzz=1.0truecm \font\tenmsb=msbm10 \font\sevenmsb=msbm7
\font\fivemsb=msbm5
\newfam\msbfam
\textfont\msbfam=\tenmsb
\scriptfont\msbfam=\sevenmsb
\scriptscriptfont\msbfam=\fivemsb

\def\xup{\overline x}

\def\zup{\overline z}
\def\ydo{\underline y}
\def\xdo{\underline x}
\def\zdo{\underline z}
\def\Ydo{\underline Y}
\def\Xup{\overline X}
\def\zut{\tilde z}
\def\prof{\vrule depth 3.5pt width 0pt}
\def\sti#1#2#3{\prof{\smash{\hbox{\vtop{\hbox{$#1$}%
\hbox{\raise#2pt\hbox{$\mkern#3mu\tilde{}$}}}}}}}
\def\zdt{\sti z 9 4}
\null \bigskip

\centerline{\bf The hunting for the discrete Painlev\'e VI equation is over}

\vskip 2truecm
\bigskip
\centerline{\scap B. Grammaticos}
\centerline{\sl GMPIB , Universit\'e Paris VII} \centerline{\sl Tour 24-14,
5${}^e$\'etage} \centerline{\sl 75251 Paris, France}
\bigskip
\centerline{\scap A. Ramani}
\centerline{\sl CPT, Ecole Polytechnique} \centerline{\sl CNRS, UMR 7644}
\centerline{\sl 91128 Palaiseau, France}

\vskip 2truecm Abstract \medskip 
We present the discrete, $q$-, form of the Painlev\'e VI equation written as a three-point
mapping and analyse the structure of its singularities. This discrete equation goes over to 
P$_{\rm VI}$ at the continuous limit and degenerates  towards the discrete $q$-P$_{\rm V}$
through coalescence. It possesses special solutions in terms of the $q$-hypergeometric
function. It can bilinearised and, under the appropriate assumptions, ultradiscretised. A new
discrete form for P$_{\rm V}$ is also obtained which is of difference type, in contrast with
the `standard' form of the discrete P$_{\rm V}$. Finally, we present the `asymmetric' form of
$q$-P$_{\rm VI}$ as a system of two first-order mappings involving seven arbitrary parameters.

\vfill\eject

\footline={\hfill\folio} \pageno=2

\bigskip

The six Painlev\'e equations were discovered a century ago by Painlev\'e [1] who responded to
the challenge of Picard [2] by classifying the integrable second-order equations of the form
$w''=f(w',w,t)$ and discovered the six transcendental equations that were subsequently named
after him. These equations define new functions which extend into the nonlinear domain the
special functions of the hypergeometric family, while constituting the nonautonomous extensions
of the elliptic functions. Discrete analogues of these equations have been proposed much later,
although, already at the time of Painlev\'e, Laguerre [3] had examined integrable nonautonomous
discrete systems and Shohat [4], half a century later, had proposed an equation which is known
today as the discrete P$_{\rm I}$. Discrete Painlev\'e equations  were subsequently derived in
various contexts [5,6] and, finally, identified as such in [7]. All these equations were of
difference type, where the independent variable $n$ enters in an additive way. Multiplicative,
$q$-discrete, Painlev\'e equations were obtained for the first time in [8] where we presented
the discrete, or $q$-discrete, forms of the first five Painlev\'e equations. Subsequent work
has led to the proposal [9] of a discrete analogue for all (but one) equations of the
Painlev\'e-Gambier [10] classification.

Still, one element was missing: the discrete form of Painlev\'e VI. A first step in this direction
was accomplished when Jimbo and Sakai [11] discovered that the `asymmetric', two-component, form of
$q$-P$_{\rm III}$ has indeed P$_{\rm VI}$ as its continuous limit. Shortly afterwards the asymmetric
forms of d-P$_{\rm IV}$ and $q$-P$_{\rm V}$ were also shown [12] to go over to P$_{\rm VI}$ at the
continuous limit. (We must point out here that the latter discrete equation possesses more
parameters than the asymmetric $q$-P$_{\rm III}$ and thus, {\sl as a discrete system}, goes beyond
P$_{\rm VI}$). Despite these interesting results, the question of the existence of a
one-component, `symmetric' in the QRT terminology [13], form for the discrete P$_{\rm VI}$
remained open. In previous papers we have speculated on the form of this equation [14] and its
singularity structure [9], but its precise form was missing. This is remedied in this paper.

We shall start by presenting the form of the discrete P$_{\rm VI}$ which is of multiplicative,
$q$-, type. Consequently we shall refer to it as $q$-P$_{\rm VI}$. Once the form is given we
shall verify that it has all the required properties: a continuous limit to P$_{\rm VI}$,
special solutions of $q$-hypergeometric type, a degeneration through coalescence to $q$-P$_{\rm
V}$ and, foremost, a generic, symmetric, singularity pattern (as expected from our work in
[9]). The bilinearisation of $q$-P$_{\rm VI}$, its generalisation to as asymmetric form and its
ultradiscretisation [15] complete this paper. One caveat is in order at this point. No
systematic derivation of $q$-P$_{\rm VI}$ will be presented. Its form was not derived from
first principles but, rather, through intuition and inspiration, two invaluable investigation
ingredients.

Let us give the form of $q$-P$_{\rm VI}$. With the usual notation $x\equiv x(n)$, $\xup\equiv
x(n+1)$, $\xdo\equiv x(n-1)$, we have:
$${(x\xup-z\zup)(x\xdo-z\zdo)\over (x\xup-1)(x\xdo-1)}=
{(x-az)(x-z/a)(x-bz)(x-z/b)\over(x-c)(x-1/c)(x-d)(x-1/d)}\eqno(1)$$
where $z=z_0\lambda^n$ (and, of course, $\zup=\lambda z$, $\zdo=z/\lambda$) and $a,b,c,d$ are
free constants. The singularity pattern of equation (1) can be obtained in a straightforward
way. First, suppose that at a given iteration we have $x=c$ while $\xdo$ has a generic value.
This means that the rhs of (1) diverges. This has as a consequence $x\xup=1$ and thus
$\xup=1/c$ (i.e. a vanishing denominator at the next step), leading to regular values for $x$
thereafter. Similarly, if $x$ goes through a zero of the numerator, $x=az$ for instance, we
must have $x\xup=z\zup$ and thus $\xup=\zup/a$. Thus the numerator will vanish at the next
step ensuring the confinement [16] of this singularity. Finally, from the form of (1), it is
clear that $x$ can go through the value zero. A detailed analysis of this situation leads to
the condition $\zup\zdo=z^2$, in which case $x=0$ is not a singularity at all. Thus we find
$z=z_0\lambda^n$. The net result of this analysis is that the singularities of $q$-P$_{\rm VI}$ are
grouped in pairs leading to 8 singularity patterns: $\{az,z/a\}$, $\{z/a,az\}$, $\{bz,z/b\}$,
$\{z/b,bz\}$, $\{c,1/c\}$, $\{1/c,c\}$, $\{d,1/d\}$, $\{1/d,d\}$ where confinement occurs in just one
step. This is precisely what we have postulated in [9] where we stated that
$q$-P$_{\rm VI}$, just as its continuous counterpart should possess a symmetric and generic
singularity pattern.

Equation (1) is indeed a discrete form of Painlev\'e VI as can be assessed from the continuous
limit. Putting $\lambda=e^\epsilon$, $a=-e^{\epsilon\alpha}$, $b=e^{\epsilon\beta}$,
$c=-e^{\epsilon\gamma}$, $d=e^{\epsilon\delta}$ we find in the limit $\epsilon\to0$:
$$\displaylines{{d^2 x\over dz^2}={1\over 2}\left({1\over x+1}+{1\over x-1}+{1\over
x+z}+{1\over x-z}\right)\left({d x\over dz}\right)^2 
-\left({1\over z}+{1\over z-1}+{1\over z+1}+{1\over x-z}-{1\over x+z}\right){d x\over dz}
\hfill \cr \hfill +{(x^2-z^2)(x^2-1)\over z^2(z^2-1)}\left({(\alpha^2-1/4)z\over
(x+z)^2}-{(\beta^2-1/4)z\over (x-z)^2}-{\gamma^2\over
(x+1)^2}+{\delta^2\over(x-1)^2}\right)\quad(2)\cr}$$
which is indeed Painlev\'e VI in non canonical form. In order to make it canonical we introduce the
following change of variables: $z=(1+\sqrt\zeta)/(1-\sqrt\zeta)$ and
$x=(\sqrt\zeta+w)/(\sqrt\zeta-w)$ and obtain:
$$\displaylines{{d^2 w\over d\zeta^2}={1\over 2}\left({1\over w}+{1\over w-1}+{1\over
w-\zeta}\right)\left({d w\over d\zeta}\right)^2-\left({1\over \zeta}+{1\over \zeta-1}+{1\over
w-\zeta}\right){d w\over d\zeta}
\hfill \cr \hfill +{w(w-1)(w-\zeta)\over 2\zeta^2(\zeta-1)^2}\left(\gamma^2-{\delta^2\zeta\over
w^2}+{\alpha^2(\zeta-1)\over (w-1)^2}+{(1-\beta^2)\zeta(\zeta-1)\over
(w-\zeta)^2}\right).\quad(3)\cr}$$

The discrete P$_{\rm VI}$ equation falls nicely into the pattern of degeneration through coalescence
that has been established for the continuous Painlev\'e equations and which has been
verified for the discrete Painlev\'e as well [8,17]. Indeed, the proper limit of $q$-P$_{\rm
VI}$ allows us to recover $q$-P$_{\rm V}$. Following our customary notation [18], we shall
denote by uppercase letters the variables of $q$-P$_{\rm VI}$ and by lowercase those of
$q$-P$_{\rm V}$. We put $X=x$, $Z=z/\delta$, $A=a/\delta$, $B=b/\delta$, $C=c$, $D=d$ and then
take the limit $\delta\to0$. We obtain thus:
$$(x\xup-1)(x\xdo-1)={(x-c)(x-1/c)(x-d)(x-1/d)\over(1-ax/z)(1-bx/z)}\eqno(4)$$
which is precisely $q$-P$_{\rm V}$ in canonical form.

The continuous P$_{\rm VI}$  equation has solutions in terms of hypergeometric functions for
some special values of the parameters. The same holds true for $q$-P$_{\rm VI}$. The simplest
way to obtain these special solutions is to use the splitting technique [19]. We separate
equation (1) in two discrete Riccati, homographic, equations in the following way:
$${x\xup-z\zup\over x\xup-1}={(x-az)(x-bz)\over(x-c)(x-d)}\eqno(5a)$$
$${x\xdo-z\zdo\over x\xdo-1}={(x-z/a)(x-z/b)\over(x-1/c)(x-1/d)}\eqno(5b)$$
The two equations of system (5) are indeed homographic and compatible provided the condition
$ab=\lambda cd$ holds. The linearisation of the discrete Riccati is obtained through a Cole-Hopf
transformation
$x=P/Q$, resulting to the linear equation
$$\displaylines{\overline Q(az-d)(az-c)((a+b)\zdo-c-d)\hfill\cr\hfill
+aQ\left((a+b)\zdo((cd\lambda-1)z^2+\lambda-cd)-(c+d)((ab-1/\lambda)z^2+1-ab/\lambda)\right)
\hfill\cr\hfill-\underline Q(a-dz)(a-cz)((a+b)z-c-d)=0\quad(6)\cr}$$  
Equation (6) has the hypergeometric
equation as continuous limit. This limit is simpler to obtain if we start from the discrete Riccati
(5) and implement the continuous limit introduced above. With the same ans\"atze we are thus led to
a continuous Riccati equation which can be linearized through the Cole-Hopf
$w=\zeta-{\zeta(1-\zeta)\over\gamma u}{d u\over d\zeta}$ leading to
$$\zeta(1-\zeta){d^2 u\over d\zeta^2}+(\beta-\delta-(\beta+\gamma+1)\zeta){d u\over d\zeta}
-\beta\gamma u=0\eqno(7)$$
i.e. precisely the Gauss-hypergeometric equation. Just like continuous P$_{\rm VI}$, equation (1)
has also rational solutions. The simplest one, $x=\sqrt z$, is obtained provided the conditions $a=c$
and $b=d$ hold. Higher solutions of both linearisable and rational type can be obtained throught the
application of the Schlesinger transformations of $q$-P$_{\rm VI}$, which are currently under study
[20].

Another interesting aspect of $q$-P$_{\rm VI}$ is its bilinearisation. In [14] we have
explained how the bilinear, Hirota, form can be obtained using the singularity pattern as a
guide. Indeed, we expect the number of $\tau$-functions to be equal to the different
singularity patterns [21]. We start by introducing 8 $\tau$-functions $\tau_a$, $\tau_b$,
$\tau_c$, $\tau_d$, $\sigma_a$, $\sigma_b$, $\sigma_c$, $\sigma_d$ and regroup them in
combinations $\phi_1=\tau_a\overline\sigma_a$, $\phi_2=\overline\tau_a\sigma_a$,
$\phi_3=\tau_b\overline\sigma_b$, etc. Parallely we introduce the 8 quantities $\alpha_1=z/a$,
$\alpha_2=za$, $\alpha_3=z/b$, etc. The bilinearisation is introduced by the following ansatz for
$x$: $x=\alpha_i-\phi_i/\psi$, 
$i=1,\dots,8$ where $\psi$ is, a priori, bilinear in the $\tau$-functions $\tau, \sigma$.
From the 8 possible definitions of $x$ we obtain, after elimination of $\psi$, 6
independent bilinear equations. We can write them in a most symmetric form as:
$$\phi_i(\alpha_j-\alpha_k)+\phi_j(\alpha_k-\alpha_i)+\phi_k(\alpha_i-\alpha_j)=0\eqno(8)$$
where we have used the fact that $\psi=(\phi_i-\phi_j)/(\alpha_i-\alpha_j)$ for all $i\ne j$. 
There are 56 such equations which can indeed be obtained from 6 independent ones, for instance those
with $i=1$, $j=2$ and $k=3,\dots,8$. The remaining
two bilinear equations can be obtained by substituting the ansatz for $x$ into (1). The net
result, after all possible simplifications, is an equation that can be separated into two
bilinear ones:
$$a\overline\sigma_a\underline\tau_a-{1\over a}\underline\sigma_a\overline\tau_a
=K(a-{1\over a})\sigma_b\tau_b\eqno(9a)$$
$$a\overline\sigma_c\underline\tau_c-{1\over c}\underline\sigma_c\overline\tau_c
=K(c-{1\over c})\sigma_d\tau_d\eqno(9b)$$
Equation (9a) can be written in a different way where $a$ and $b$ are permuted, but this two forms
are equivalent, provided equations (8) hold. The same holds true for equation (9b) under a
permutation of $c$ and $d$. It is clear from the form of equations (9) that
the value of $K$ can be modified through a gauge transformation which leaves (8) invariant. Thus,
with the appropriate choice of gauge, we can take $K=1$ without loss of generality. The system that
consists of equations (9) and 6 independent equations among those of (8) is the bilinear
form of $q$-P$_{\rm VI}$.

One final point concerning $q$-P$_{\rm VI}$ is its ultradiscretisation. This method has been
introduced by the Tokyo-Kyoto group [15] as a method for obtaining discrete equations where,
provided one ensures the right parameters and initial conditions, the dependent variable takes
only integer values. Thus the evolution introduces a generalised cellular automaton. The
ultradiscretisation of $q$-P$_{\rm VI}$ is obtained using the standard techniques of [22]. We
first rewrite (1) as:
$$y\ydo={(x+az)(x+z/a)(x+bz)(x+z/b)\over(x+c)(x+1/c)(x+d)(x+1/d)}\eqno(10a)$$
$$x\xup={y+z\zup\over y+1}\eqno(10b)$$
by inverting the sign of the constants $a,b,c,d$. Next, we introduce
$\lambda=e^{1/\epsilon}$, $x=e^{X/\epsilon}$,
$y=e^{Y/\epsilon}$ and similar ans\"atze for $z,a,b,c,d$. The limit of (10) which uses
extensively the identity
$\lim_{\epsilon\to0}\log(e^{\alpha/\epsilon}+e^{\beta/\epsilon})=\max(\alpha,\beta)$, leads to
the ud-P$_{\rm VI}$:
$$\displaylines{Y+\Ydo=\max(X,Z+A)+\max(X,Z-A)+\max(X,Z+B)+\max(X,Z-B)\hfill\cr\hfill
-\max(X,C)-\max(X,-C)-\max(X,D)-\max(X,-D)\quad(11a)\cr}$$
$$X+\Xup=\max(Y,2Z+1)-\max(Y,0)\eqno(11b)$$
The study of P$_{\rm VI}$ leads naturally to more equations that we will briefly examine
here. The first is a `symmetric', one component, difference equation which can be obtained almost by
inspection from $q$-P$_{\rm VI}$, by considering a different coalescence. Indeed taking $X=1+\delta
x$, $\lambda=1+\delta\kappa$, $A=1+\delta a$ etc., we find:
$${(x+\xup-z-\zup)(x+\xdo-z-\zdo)\over (x+\xup)(x+\xdo)}=
{(x-z-a)(x-z+a)(x-z-b)(x-z+b)\over(x-c)(x+c)(x-d)(x+d)}\eqno(12)$$
where $z=z_0+\kappa n$. This equation is a new discrete form of P$_{\rm V}$. This can be
assessed from the continuous limit of (12). Putting $z=\sqrt\zeta$, $x=\sqrt\zeta/(1-w)$ and 
$a=\alpha\epsilon^2$, $b=\beta/\epsilon^2$, $c=\gamma\epsilon^2$, $d=\beta/\epsilon^2+\delta$ we
obtain:
$${d^2 w\over d\zeta^2}=\left({1\over 2w}+{1\over w-1}\right)\left({d w\over
d\zeta}\right)^2-{1\over \zeta}{d w\over d\zeta}
+{(w-1)^2\over 2\zeta^2}(\alpha w-{\gamma\over w})+{\delta w\over 2\beta\zeta}-
{w(w-1)\over 2\beta^2(w-1)}.\eqno(13)$$

Finally we can ask whether there exist asymmetric (two-component) extensions of both (1) and
(12). It turns out that this is straightforward. We obtain thus the 
asymmetric $q$-P$_{\rm VI}$
$${(y\xup-z\zut)(xy-z\zdt)\over (y\xup-1)(xy-1)}=
{(y-az)(y-bz)(y-cz)(y-dz)\over(y-p)(y-q)(y-r)(y-s)}\eqno(14a)$$
$${(xy-z\zdt)(x\ydo-\zdt\zdo)\over (xy-1)(x\ydo-1)}=
{(x-\zdt/a)(x-\zdt/b)(x-\zdt/c)(x-\zdt/d)\over(x-1/p)(x-1/q)(x-1/r)(x-1/s)}\eqno(14b)$$
where $\zut=z\sqrt\lambda$, $\zdt=z/\sqrt\lambda$, and with the constraints $abcd=1$ and $pqrs=1$.
Similarly, we find the asymmetric d-P$_{\rm V}$
$${(y+\xup-z-\zut)(x+y-z-\zdt)\over (y+\xup)(x+y)}=
{(y-z-a)(y-z-b)(y-z-c)(y-z-d)\over(y-p)(y-q)(y-r)(y-s)}\eqno(15a)$$
$${(x+y-z-\zdt)(x+\ydo-\zdt-\zdo)\over (x+y)(x+\ydo)}=
{(x-\zdt+a)(x-\zdt+b)(x-\zdt+c)(x-\zdt+d)\over(x+p)(x+q)(x+r)(x+s)}\eqno(15b)$$
where $\zut=z+\kappa/2$, $\zdt=z-\kappa/2$, and with the constraints $a+b+c+d=0$ and
$p+q+r+s=0$. These equations are the most general discrete Painlev\'e equations known to date. They
have seven discrete parameters and thus they lie beyond the already known asymmetric $q$-P$_{\rm
III}$, d-P$_{\rm IV}$ and $q$-P$_{\rm V}$. Still, we expect their continuous limits to go over to
P$_{\rm VI}$. The fact that these equations do not have the constraints of the symmetrical form make
us believe that they possess the property of self-duality just like asymmetric $q$-P$_{\rm III}$ [23]
and asymmetric d-P$_{\rm IV}$ and $q$-P$_{\rm V}$ [24]. Thus it is interesting to study these
equations in the framework of what we call the Grand Scheme i.e. the simultaneous analysis of a
discrete equation and its Schlesinger transformations. This aspect of $q$-P$_{\rm VI}$ and 
d-P$_{\rm V}$ is the object of a future study [20].

\bigskip {\scap References}.
\medskip
\item{[1]} P. Painlev\'e, Acta Math. 25 (1902) 1.
\item{[2]} E. Picard, Journal de Liouville 5 (1889) 135.
\item{[3]} E. Laguerre, J. Math. Pures Appl. 1 (1885) 135.
\item{[4]} J.A. Shohat, Duke Math. J. 5 (1939) 401
\item{[5]} M. Jimbo and T. Miwa, Proc. Japan Acad. 56 A (1980) 405.
\item{[6]} M. Jimbo and T. Miwa, Physica 2D (1981) 407. 
\item{[7]} E. Br\'ezin and V.A. Kazakov, Phys. Lett. 236B (1990) 144.
\item{[8]} A. Ramani, B. Grammaticos and J. Hietarinta, Phys. Rev. Lett. 67 (1991) 1829.
\item{[9]} B. Grammaticos, A. Ramani, Meth. and Appl. of An. 4 (1997) 196.
\item{[10]} B. Gambier, Acta Math. 33 (1910) 1.
\item{[11]} M. Jimbo and H. Sakai, Lett. Math. Phys. 38 (1996) 145. 
\item{[12]} A. Ramani, B. Grammaticos, Y. Ohta, CRM Proceedings and Lecture Notes v. 9, Eds. D.
Levi, L. Vinet and P. Winternitz, AMS (1996) 303.
\item{[13]} G.R.W. Quispel, J.A.G. Roberts and C.J. Thompson, Physica D34 (1989) 183. 
\item{[14]} A. Ramani, B. Grammaticos and J. Satsuma, J. Phys. A 28 (1995) 4655.
\item{[15]} T. Tokihiro, D. Takahashi, J. Matsukidaira and J. Satsuma, Phys. Rev. Lett. 
76 (1996) 3247.
\item{[16]} B. Grammaticos, A. Ramani and V. Papageorgiou, Phys. Rev. Lett. 67 (1991) 1825.
\item{[17]} B. Grammaticos, Y. Ohta, A. Ramani, H. Sakai, J. Phys. A 31 (1998) 3545.
\item{[18]} A. Ramani and B. Grammaticos, Physica A 228 (1996) 160.
\item{[19]} B. Grammaticos, F.W. Nijhoff, V.G. Papageorgiou, A. Ramani and J. Satsuma, Phys.
Lett. A185 (1994) 446. 
\item{[20]} A. Ramani and B. Grammaticos, in preparation.
\item{[21]} Y. Ohta, A. Ramani, B. Grammaticos, K.M. Tamizhmani, Phys. Lett. A 216 (1996) 255.
\item{[22]} B. Grammaticos, Y. Ohta, A. Ramani,  D. Takahashi, K.M. Tamizhmani, Phys. Lett. A 
226 (1997) 53.
\item{[23]} A. Ramani, Y. Ohta, J. Satsuma and B. Grammaticos, Comm. Math. Phys. 192 (1998) 67.
\item{[24]} A. Ramani, Y. Ohta, and B. Grammaticos, in preparation.
\end